\documentclass[conference]{IEEEtran}
\IEEEoverridecommandlockouts

\usepackage{cite}
\usepackage{amsmath,amssymb,amsfonts}
\usepackage{algorithmic}
\usepackage{graphicx}
\usepackage[numbers]{natbib}
\usepackage{textcomp}
\usepackage{xcolor}
\usepackage[normalem]{ulem}
\useunder{\uline}{\ul}{}
\def\BibTeX{{\rm B\kern-.05em{\sc i\kern-.025em b}\kern-.08em
 T\kern-.1667em\lower.7ex\hbox{E}\kern-.125emX}}
\usepackage{float}

\begin{document}

\title{Correlation of biological and computer viruses through evolutionary game theory}

\author{\IEEEauthorblockN{Dimitris Kostadimas, Kalliopi Kastampolidou and Theodore Andronikos}
\IEEEauthorblockA{\textit{Ionian University} \\
\textit{Department of Informatics}\\
Corfu, Greece \\
Email: \{p19kost2, c17kast, andronikos\}@ionio.gr}}

\maketitle

\begin{abstract}
	Computer viruses have many similarities to biological viruses, and their association may offer new perspectives and new opportunities in the effort to tackle and even eradicate them. Evolutionary game theory has been established as a useful tool for modeling viral behaviors. This work attempts to correlate a well-known virus, namely Virlock, with the bacteriophage $\phi6$. Furthermore, the paper suggests certain efficient strategies and practical ways that may reduce infection by Virlock and similar such viruses.
\end{abstract}


\section{Introduction}


\subsection{Computer viruses}

The most common term that is currently being used to refer to malicious computer programs is that of \emph{computer virus}. The rationale behind this metaphorical term being used to describe this kind of software is based in its observed behavior that appears to have very much in common with the behavior of biological viruses. Fred Cohen, the American computer scientist who coined this term in 1983 described them as self-replicating programs that infect other ones by embedding their ill-intended code in them \cite{Cohen1987}. 

Computer viruses, just like their biological counterparts, appear in many different variations. Traits like their operational conditions, their host demographic target, their replication manners, the infection type, the infection success rate, the way they spread, the consequences and their severity, and many other traits are what differentiate each and every computer virus while many are named after these kinds of traits (like Ransomware for example). Viruses are also usually classified based on the way they infect the host machine \cite{Kaspersky2021a}, \cite{uniserve}.

In the past, a computer virus was created for fun, leaving no permanent damage to the host computer. However, this has changed. A common incentive behind computer viruses is to bring profit to the owner of the malicious software, or generally benefit him in some way. One of the ways this is achieved is by stealing the infected host’s private data, encrypting them and asking the victim (also known as the sucker) to pay a ransom in exchange for the key to decrypt the encrypted data. It appears that recently there is an increase in popularity of viruses that attack the privacy of computer users in order to bring profit (illegally) to certain individuals. This calls for more ways and layers of protection to be established. 

As mentioned before, today’s computer viruses come with numerous traits and are usually categorized based on their infection mechanics and behavior. Viruses are part of the \emph{malware} family, just like \emph{worms} are \cite{Latto2021}. The term \emph{malware} refers to any kind of malicious software. Some common types of malware are \emph{ransomware, spyware, adware, trojan horses, rootkits, keyloggers}, and of course more general types like viruses and worms \cite{crowdstrike.com2021}, \cite{Kaspersky2021b}. It is worth noting that there are many more types of malware.
For instance, ransomware is a type of computer virus that encrypts the victim’s data and asks them to pay a ransom in exchange for the decryption key or general access back to the computer in case of a total lockdown \cite{Wikipedia2021}.

Another trait that differentiates viruses is whether they contain/implement a worm component in them or not.
The worm component of a virus is what enables them to spread through a network of computers, leading to reproduction of its kind while also causing mutations to take place in the cases of polymorphic or metamorphic code.

For a computer virus to engage for the first time with the host computer and to start its replication, it must first infect the targeted computer. There appears to be a certain amount of randomness in the way the virus continues to spread in the cases of worms and viruses with a worm component and with a polymorphic or metamorphic code. There are many ways for a malware to infect a system. The most common ways are through email attachments, downloading suspicious files, use of non-trustworthy removable media, security vulnerabilities, P2P file sharing, malvertising (that is ads that promote malicious software, found even in trustworthy sites), as well as through the network due to the worm component of a virus (that got spread from another infected computer) without the users taking any actions \cite{Latto2021}.

A usual virus would replicate itself by attaching its malicious code to multiple host computer files as the time passes. Worms on the other hand do not even need to manipulate the files in a computer in order to duplicate themselves \cite{Times2021}. Worms remain active in the memory and the CPU of the computer and their actions are usually invisible to the user except when they consume enormous amounts of the computer’s resources, in which case the slow performance will hint their existence.
The most interesting thing regarding worms and the viruses that contain a worm component, is not only that they are able to replicate themselves in a rapid way, but mainly the fact that they can do that without any human interaction with the computer whatsoever \cite{Vick2017}, \cite{Latto2021}, \cite{Norton2019}.

In addition to replication manners and dynamics, some specific types of viruses have the ability to mutate in order to achieve better success rates, better spread, and generally enhance their already existing traits. This ability is a trait of \emph{polymorphic} and \emph{metamorphic} viruses. More specifically, a \emph{polymorphic} virus makes use of a variable encryption to encrypt itself in order to make every copy of itself unique.

Another way virus mutants could appear is when the creator of the malicious software has awareness of the virus’s state in the affected computers and its effectiveness, and wants to make the malware more powerful and spread some new variation instead. The creator might take the decision to alter the original code of the virus and start spreading the new and updated mutant of the original virus in order to make it more effective and accomplish ill-intended goals.

The intent behind the use of this kind of polymorphic and metamorphic code is to evade their detection from antimalware and antivirus software. Implementing the ``polymorphic trait'' into a virus tends to be a fairly easier in comparison to the ``metamorphic trait.'' However, the cost of the implementation might be worth the extra effort as it also offers better protection against antivirus software because it renders the virus way harder to be detected. In order to prevent detection, some worms and viruses, especially the ones with a worm component, manage to implement stealth strategies. 
Some hide themselves by not taking up more space when replicating themselves by getting attached into the host’s files. Other viruses attempt to kill processes run by active antivirus software in the computer or the operating system to protect themselves and let the user have a false sense of security.

\subsection{Biological viruses}

Biological viruses are organisms that act parasitically and need to infect a host in order to be able to reproduce and carry their genetic material, either DNA or RNA and proteins. They cannot synthesize proteins and, therefore, use host ribosomes to translate their RNA into proteins that serve them. Viruses are transmitted in different ways, depending on their species. The number of cells infected with a virus is called ``host range.'' The most prominent way of dealing with a biological virus is the immune system of the organism whose body it will infect. Usually the infected organisms are animals, plants, molecules, and, of course, humans. Additionally, vaccines provide a good defense and help the immune system, usually in regard to a specific virus infection. Apart from vaccines, antiviral drugs are also available and are evolving over time. However, there are some categories of viruses that attack the organism’s immune system that they have infected, which cause chronic infections.

When a cell is infected with a virus, it necessarily and directly replicates itself in copies of the virus. Viruses are made up of their genetic material, the capsid, a set of proteins that protect the genetic material, and in some cases from external lipids. The virion is the extracellular form of the virus. Depending on their genome, whether it is a DNA or an RNA genome, they are called as such (DNA \& RNA virus respectively). For an RNA virus, the genetic material consists of ribonucleic acid (RNA) \cite{Wagner2004}.

The effects that a virus has on an organism are numerous. Most cause the death of the host cell. Usually, death involves restricting the normal activity of the cell by viral proteins. The effects of some viruses can cause permanent damage to the host organism or can be eliminated without malignancy. Some viruses infect an organism and do not cause changes in the cells. Therefore, their function continues normally with the virus, however ending up infecting persistently eventually. Virome is the set of viruses that infects an organism. Phage typing is a common method for tracing the source of infections \cite{baggesen2010phage}.

Precisely because viruses are acellular organisms, they are not transmitted by cell division. For this purpose, they use the host, in order to create many copies of themselves. When a virus infects a host, the host is forced to reproduce the original virus. There is a basic life cycle for viruses. Infection begins with the attachment of a virus and its proteins to the surface of the host. At this point, the host range and cell type are determined. This is followed by the penetration of virions into the cell. Bacteria do not have a strong protective wall and viruses have developed mechanisms for gene penetration, while the capsid remains outside the cell. The final step is the release of the virus into the host cell by the Uncoating process \cite{blaas2016viral}.

Replicating the virus also means multiplying the genome. After replication, particles and altered proteins may appear relative to the original form of the virus prior to penetration. Lysis is the process by which a virus is released from a host cell. This causes the cell to be killed. Prophage is the process by which the host reproduces, so the virus is also replicated. When the virus ceases to be inactive, it causes a lysis in the host cell. Reproduction of an RNA virus occurs in the cytoplasm. Each virus has its own enzymes that make copies of the genomes. After lysis, the virus can infect another, new host cell, leading to the repetition of this cycle. Also, during this step, there may be mutations of the virus \cite{rogers2010bacteria}. When an organism's immune system is exposed to a virus, it produces antibodies to suppress the virus. This process is called humoral immunity. Depending on the antibodies that are produced, it remains to be seen whether the body has recovered from the virus or not.

Bacteriophages or phages are viruses that alter or diverse microbial populations. They were used as antibacterial agents due to their properties \cite{onodera1992construction}. The host range of some bacteriophages focuses on a single bacterial strain. Bacteriophages are one of the groups of viruses and infect specific bacteria. They usually have double-stranded RNA genomes.

An RNA virus consists of segments that make up a protein. These segments exist in the capsid. Different segments may be in different virions and yet the virus may be contagious. The way they do the infection is by attaching themselves to molecules of the surface of the bacterion and then they enter the cell. In many cases, as soon as the original virus enters the cell, it begins to translate its mRNA into proteins. Then, the result of this process either becomes virion and helps in the formation of other such virions or participates in the process of cell lysis. Virus enzymes contribute to the destruction of the cell membrane. The main way in which bacteria are protected from such infections is with the help of enzymes that target unknown RNA. Bacteria can also detect genomes from viruses that have fought in the past and block their reproduction through RNA interference. This is a mechanism of defense for bacteria against such infections. By their nature, bacteria have the tactic of interfering RNA. During the replication of a viral RNA, some mutations occur, which may either not affect the cell proteins or contribute to a resistance to antiviral drugs.

\subsection{Evolutionary game theory}

Evolutionary game theory (EGT) is a tool that helps with the modeling of the behavior of this kind of viruses, thus setting a path on designing a higher level of security. It has captured the interest of scientists such as biologists, mathematicians, economists, psychologist, computer scientists and many more. Its connections and applications to realistic real-life conditions is what makes this concept even more captivating. EGT provides scientists with the appropriate tools to study instinctive behaviors, biological phenomena and even decisions based on rationality. The term population dynamics is now widely used and refers to phenomena and behaviors just like the above, as well as sets of strategies or characteristics players might inherit \cite{weibull1997evolutionary}.

The microscopic level is just as interesting as the macroscopic kind of observation. Even in biological processes, game properties have been observed. Taking a deeper look into the cells and macromolecules that are part of multicellular organisms, game properties can be observed. The strategy that each one of them adopts is based on their moves. The strategy of a certain player can (and probably will) change as they follow the principle of natural selection. Strategies are always subject to change as throughout the cell’s lifespan mutations tend to happen which cause irreversible changes, as well as reversible changes caused by epigenetic modifications. All the above are related to the reproduction of these objects and it is evident that reproductive success alters the game’s outcome.

A brief overview of evolutionary games in the context of biological systems is given in \cite{Kastampolidou2020}. The incorporation of games in biology is a bigger persistent trend. Many classical games, including the famous Prisoners' Dilemma (see \cite{Kastampolidou2020a} for references), have been used to model biological situations. This is not limited to viruses, as it extends to microbes and their games (see \cite{Kastampolidou2020b}), and even to bio-inspired models of computation (see \cite{Giannakis2015a} and \cite{Theocharopoulou2019} for details). The introduction of unconventional tools, such as games, automata, notions from quantum mechanics, promises to bring new perspectives and new insights to the study of biological processes. For example, the adoption of concepts from game theory to the field of quantum computation has proved to be extremely successful (see \cite{Giannakis2015b}, \cite{Andronikos2018}, \cite{Giannakis2019} and \cite{Andronikos2021} for some recent results and more related references). It is worth pointing out that games may tackle critical problems; coin tossing plays a crucial role in the design of quantum cryptographic protocols (see \cite{Bennett2014} and references therein, and the more recent \cite{Ampatzis2021}).

This paper offers a new perspective on the correlation of computer viruses to biological viruses. There are of course many types of biological viruses, and the same can be said about computer viruses. The behavioral traits of biological viruses can be associated with the corresponding traits of the computer viruses. The emphasis in this work is placed on a well-known computer virus, namely Virlock, and its similarities with the biological virus $\phi6$. These similarities, along with some of the anticipated differences, are thoroughly examined and analyzed in Section \ref{sec:ComparingTheViruses}. We hope that this approach will shed new light in the adoption and application of strategies that have been successful in tackling viruses of one type to the other type, as well as enhance the means to assess the effectiveness of the employed strategies.

\section{The Virlock virus} \label{sec:VirLock}

VirLock asks its victims to pay a ransom in order to regain access to their files and their computing systems in general. VirLock has a parasitic behavior. From the time it is executed, it starts infecting the supported computer files, but the way it alters/infects a file is a bit different from what normally happens with this type of malware. VirLock embeds clean code inside a malware instead of malware inside clean code. This means that every file that has been encrypted will be embedded into the malware. Now every infected file can infect as it works like a mutation of VirLock. VirLock self-replicates itself this way and grown its ``population.''

The first detection of the VirLock virus happened in 2014 \cite{aurangzeb2017ransomware}. Of course, as it is a polymorphic virus, many different mutations have been encountered through the span of several years until today. As VirLock continues to evolve, differences were found not only in the \emph{decoration-code}, but also in its core functions. Specifically, the virus is also able to spread through networks, thanks to the cloud storage that more and more people start using nowadays.
The mutations are many and there are multiple variables of VirLock in the databases of several antivirus software and, as some results of the famous website virustotal hint, the mutations of the virus probably helped evade detection from certain antiviruses over time.

VirLock is able to occupy the whole screen area of the computer and kill the \emph{explorer.exe} task of the Windows operating system that handles the graphical user interface \cite{Sophos2016}. This means that it renders the infected computer almost useless by the time it infects it, as there is no way to access the main functions of the operating system because the whole screen is being occupied by the virus message, while binary files and files with certain extensions are being ``encrypted'' in the background. Because of this, the user has no way of using an antivirus software the conventional way. As suggested by many antivirus companies, the best way to try and disinfect a computer from VirLock is to boot into the safe mode with network capabilities that Windows OS offers. This way, VirLock probably will not be able to launch itself during the startup. If the OS is booted successfully, then the user can perform a virus scan with an antimalware or antivirus software to attempt virus detection. The chances of an antivirus software detecting the virus of course depend on what kind of VirLock variation this computer is infected with, since it is quite possible that a new variant could be unrecognizable yet. Moreover, there is a high chance that by the time the user acquires the knowledge of how to proceed in the disinfection, the virus already has encrypted most, if not all, of the computer files.

Every time VirLock encrypts a file, it appends the .exe extension to it and renders it a copy of itself.
So, every infected file is technically a variation of the malware itself. Antivirus software are usually not able to decrypt files, so the least they can do is detect the malware and quarantine it or completely delete it. Deleting all the infected files in the computer is obviously not an optimal solution. Some anti-virus software offer a VirLock cleaner that is able to wipe the virus' remnants and also ``decrypt'' most (if not all) of the infected files in case of known VirLock variants.
The user is informed that false positives may also be found and should be careful during the removal process. As VirLock has many variants, which makes it hard for antivirus software to detect it, it is clear that What would help in the detection of this kind of software is the study of its behavior.

Antivirus programs that are able to do live behavioral analysis have a definite advantage in tackling the virus. This is because the mutations in this case have something in common, and this is the core code. Even if the code mutates as it self-replicates, the core functions remain the same. By focusing on them and the way they react, it is possible to achieve a better level of protection. Even though this is important, there are still ways that VirLock dodges the emulations of antivirus software thanks to tricks like the payload encryption and its obscure code in general. Even if security is much more advanced nowadays, there’s still reportedly about 70\% of malware that manages to evade detection attempts performed by antivirus and antimalware software 
\cite{TheStateOfSecurity2015}, \cite{CyberHoot2020}.

Security specialists have also found that VirLock appears to have an exploit by itself. By entering 64 zeros in the \emph{TrasnferID} field, it is possible to trick VirLock that the ransom has been paid. After that, by clicking a file, the decrypting process is activated. The drawback of this strategy is that the user will have to do this for every single file in the computer, with the risk of infecting the computer once again. Another known exploit of this malware is that it ignores the volume shadow copies of Windows. So, if this feature is enabled and volume shadow copies of the computer are present, then the damage can be reverted. Probably the best way to protect against this kind of viruses is to hold constant backups of the precious files. Obviously, an updated antivirus software and network segmentation may also prevent the spread of the virus \cite{Orange2017}.

The latest variants of VirLock appear to be fairly powerful and effective, even though enough time has passed since its first appearance and outbreak. When VirLock infects a computer, it displays a message occupying the entire screen area informing the user that pirated software was detected in the computer. Thus, a fine must be paid for this illegal action in order to not get arrested. To persuade the user that this notice is legit, VirLock has localized GUI, so depending on the victim’s location it will display the logotypes of the corresponding local authorities and government. Meanwhile, the files of the infected computer are being encrypted in the background and are also being infected. The term \emph{encryption} in this case deserves some clarification \cite{Sjouwerman2016}.

VirLock does not use a one-way encryption algorithm like AES or RSA that some of the more popular ransomware tend to use. Instead, it performs a two-stage encryption, making use of the XOR and XOR-ROL operations \cite{Sophos2016}. Entropy is not as high as if AES and RSA were used. This operation that makes the data appear more obscure will still be referred to as encryption in the article. The infection happens by trying to run the malicious file. When VirLock is executed, it drops 3 randomly named executables in randomly named folders. As the virus is polymorphic, these executables have identical hashes that are different every time \cite{Netskope2017} and older variants appear to drop only 2 of them. One of those registers itself as a Windows service to cover itself up, while the others encrypt and infect the computer’s files. Other measurements that VirLock takes to protect itself is that it disables the task manager process so that the user is unable to take control of what is happening in the computer and kill the virus that alters the Windows Registry. These are considered to be VirLock's trademark attributes. The first registry entry that VirLock alters concerns the User Access Control (UAC). By disabling UAC, the virus is able to manipulate everything in the computer freely without the need of administration privileges. It then hides the known file extensions in order to trick the user that the files are fine and making him unable to see the .exe extension that VirLock appends, so the user might backup those files thinking they are safe and moving them to another computer or try to run them again. The last registry change is one that makes hidden files invisible \cite{Sophos2016}.


\begin{figure}[ht]
	\centering
	\includegraphics [ scale = 0.08 ] {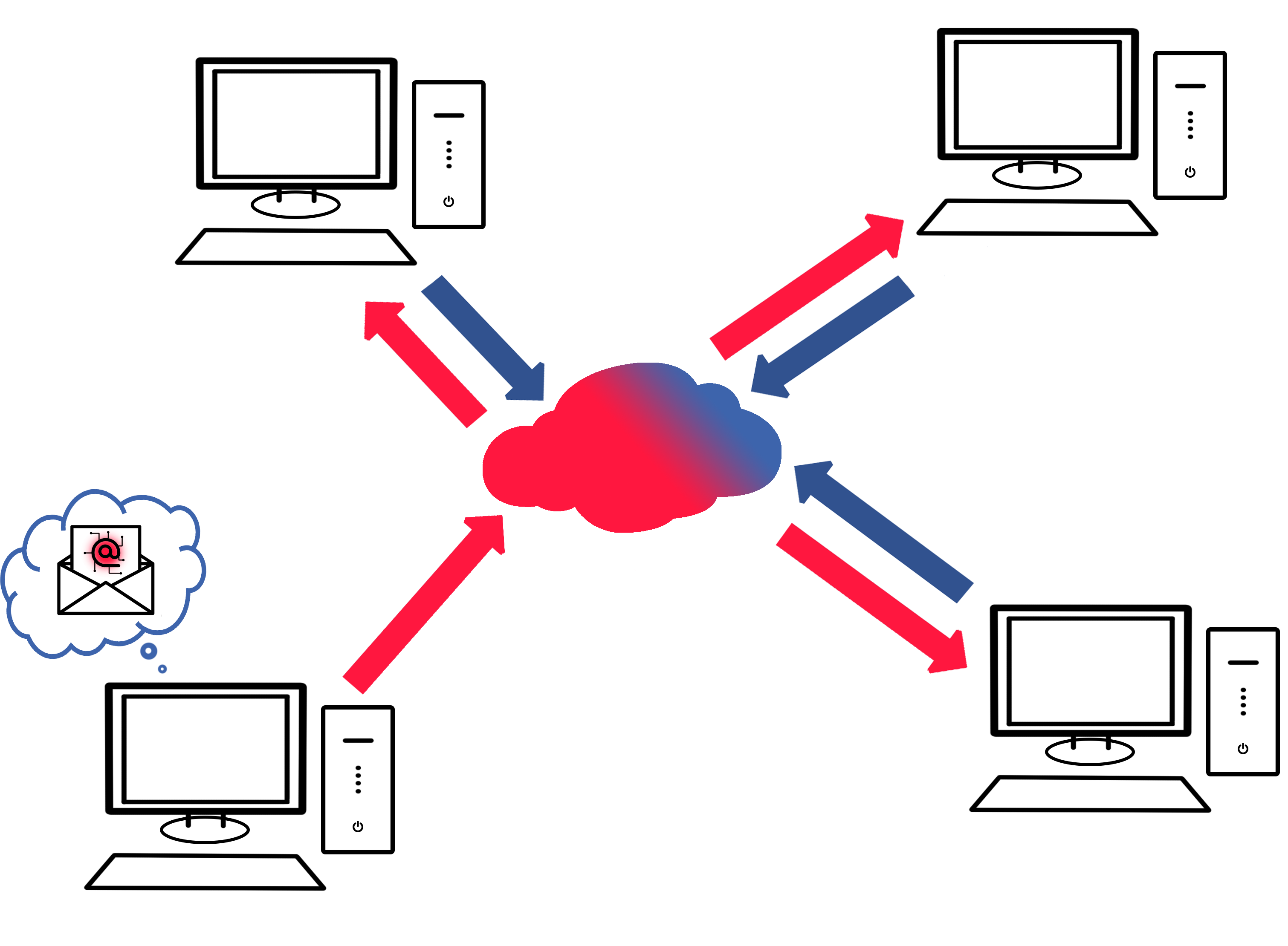}
	\caption{Representation of VirLock Cloud Storage Infection. In the case above, the bottom left computer is getting infected with the VirLock malware through a malicious email attachment that the user opened. VirLock will infect the files in the cloud storage as well, and the rest of the computers in the network might eventually get the malware. The red arrows represent the path of the infection, while the blue arrows represent interaction with the cloud.}
	\label{fig:VirLock}
\end{figure}

The structure of this type of files is as follows, going from top to down: the start and the end of the code is made out of polymorphic code that changes in every iteration, so basically the code is wrapped around from this kind of polymorphic code that we also refer to as decoration code. After the first piece of polymorphic code, the malicious code, which runs every time, appears. Right after the malicious code, VirLock embeds the clean code, which is followed by the last piece of polymorphic code \cite{Sjouwerman2016}.

Of course, there is always the possibility that the user might be tempted to pay the ransom instead of getting involved in any complicated task as they engage in the virus removal process or taking advantage of the malware exploits. There are sources claiming that a fair percentage of users proceed on paying the ransom in order to regain access to their computer and data, especially users under the age of 55.

Even though paying the ransom might be tempting for the infected users, especially for companies with data of critical importance, there is an important reason why one should think twice before proceeding with the payment. First, there is a strong ethical reason, particularly in view of the outrageous amount of money the ransomware may ask. Second, there are several sources claiming that only 8\% of those who proceed in the ransom payment manage to get the entirety of their data back. It is also reported that ``on average, only 65\% of the encrypted data was restored after the ransom was paid.'' Even if the victims are really desperate to get their data back, it appears that there are other much more effective ways to achieve that.

\section{Modeling Virlock using game theory} \label{sec:ModelingVirlockUsingGameTheory}

Modeling the above situation with the help of game theory, a clearer picture of the strategy that would benefit the user the most emerges. The following payoff matrix is constructed keeping in mind the general consensus of security specialists that ``96\% of those whose data was encrypted got their data back in the most significant ransomware attack,'' which confirms that there are other more effective ways to retrieve the data besides paying the ransom.

\begin{figure}[ht]
	\centering
	\includegraphics [ scale = 0.09 ] {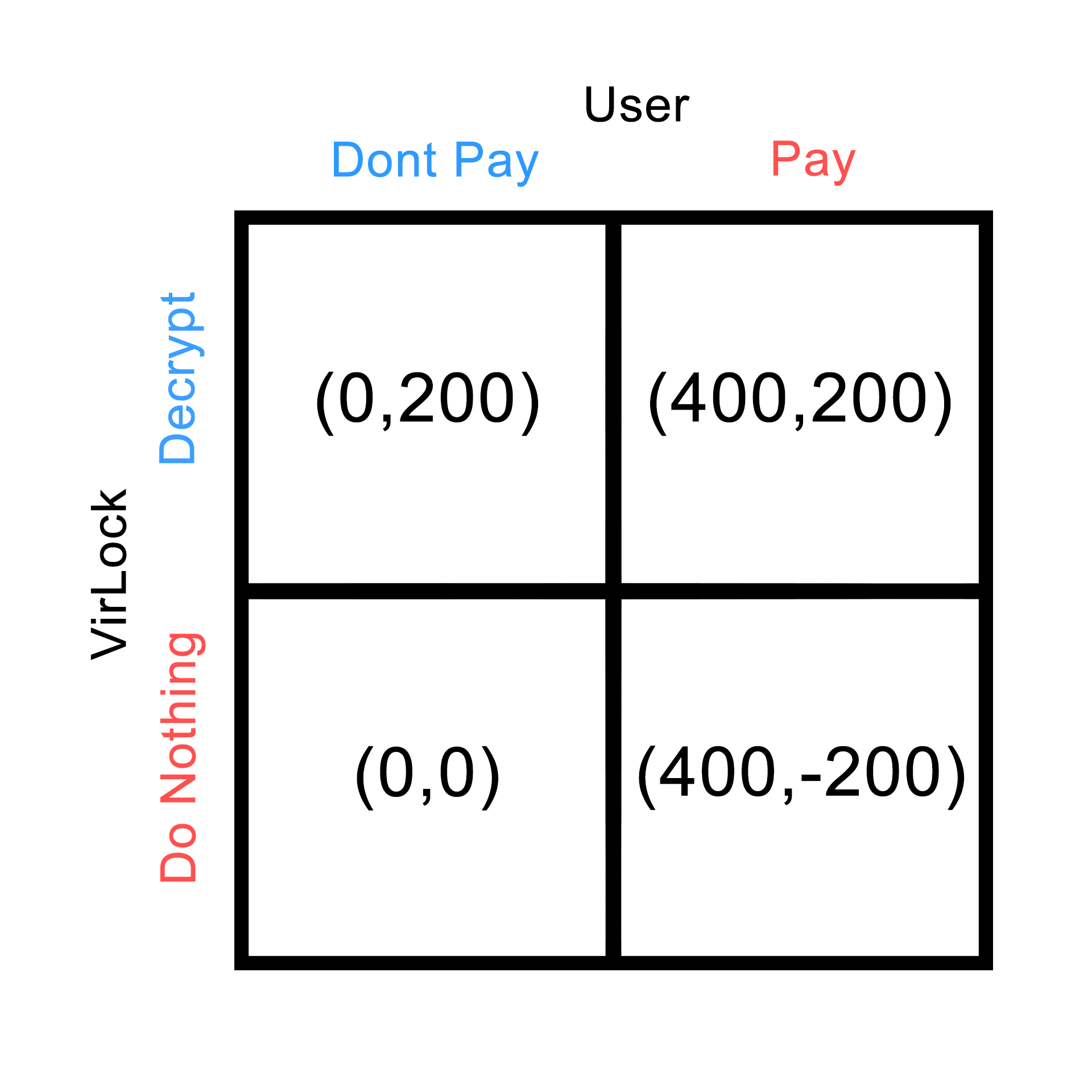}
	\caption{Ransom Payment Payoff Matrix. The user has the option to pay and the option not to pay the ransom, while the VirLock malware may or may not decrypt the users' data. The payoff matrix makes clear that the user will benefit the malware creator/s less by not proceeding in the ransom payment, and that paying the ransom holds an additional risk.
	}
	\label{fig:Payoff Matrix}
\end{figure}

\section{Pseudomonas virus $\phi6$} \label{sec:PseudomonasVirusPhi6}

Bacteriophage $\phi6$ lytic virus belongs to the cystoviridae viruses and aims to infect Pseudomonas bacteria. Its genome is double-stranded, consists of three segmented parts and codes for 12 proteins. Such species consist of a lipid membrane which surrounds their nucleocapsid.

$\phi6$ locates and then sticks to the bacterium that wants to infect with its special protein for this purpose, P3. Beyond that, different proteins contribute to the process of cell infection. The bacteriophage $\phi6$ has been widely used to model its behavior and structure, and has previously been associated with the field of classical and evolutionary game theory.

\section{Comparing the two viruses} \label{sec:ComparingTheViruses}

A comparison between the way computer viruses and biological viruses operate when they invade a host as well as their characteristics is of great interest and could potentially provide new insights. Just like their biological counterparts, computer virus types continue to evolve through the passage of time following the evolution of computers. When a computer virus mutates, generations can be observed during time-spans, just like when biological viruses mutate.

Antivirus software could metaphorically be the immune system of a computer. Computer viruses try to weaken this system in order to replicate themselves and grow their population by invading the victim’s computer files (which in this case could represent the cells of a human organism) and, as an extension, the whole network of computers connected with the original victim.

The main properties and characteristics of VirLock that could be linked to bacteriophages and especially $\phi6$ are the following:

\begin{enumerate}
	\item   They are self-replicating viruses that appear to grow exponentially.
	\item   They try to protect themselves by attacking and eventually manipulating the host.
	\item	They affect the host in order to gain full access to its functions and keep their viral ability unaffected.
	\item	They affect certain host types.
	\item	They exhibit parasitic behavior, while manipulating and embedding their code in their replicants/mutants.
	\item	They are able to spread when the infected parts come in contact with other hosts.
	\item	They have a core structure that are subject to change.
	\item	They are able to mutate rapidly not only by themselves but also with external help.
	\item	They can be untraceable for a specific time, they are, in general, hard to locate and extremely difficult to eliminate.
\end{enumerate}

\begin{figure}[H]
	\centering
	\includegraphics [ scale = 0.63 ] {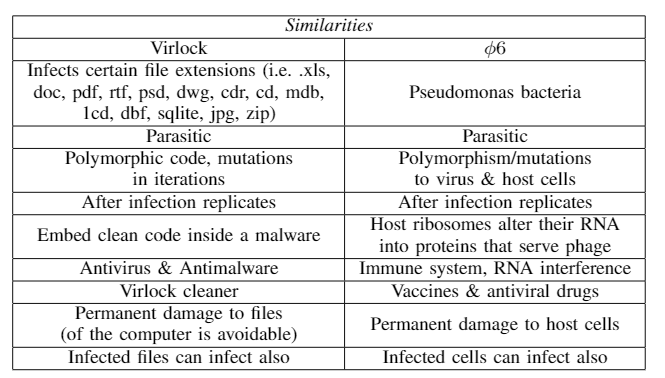}
	\caption{Similarities between the two viruses.}
	\label{fig:sim}
\end{figure}

\begin{figure}[H]
	\centering
	\includegraphics [ scale = 0.64 ] {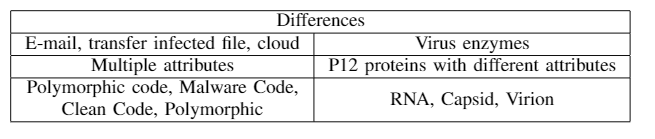}
	\caption{Differences between the two viruses.}
	\label{fig:dif}
\end{figure}

The following tables describe the steps of some of the known strategies that users could follow in order to recover their computer back to a normal functioning state. The rationale behind the following tables is to get an overall sense of the complexity inherent in every strategy, as well as the effectiveness and the risks that one has to take.

\begin{figure}[H]
	\centering
	\includegraphics [ scale = 0.7 ] {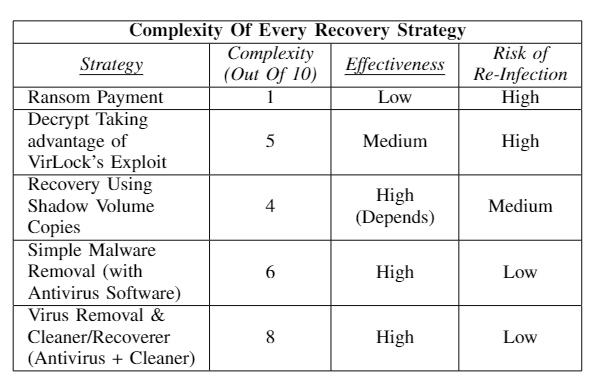}
	\caption{Complexity of the steps taken in every recovery strategy.}
	\label{fig:tab2}
\end{figure}

\begin{figure}[H]
	\centering
	\includegraphics [ scale = 1.0 ] {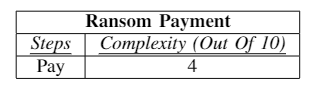}
	\caption{Ransom payment.}
	\label{fig:ransom}
\end{figure}

\begin{figure}[H]
	\centering
	\includegraphics [ scale = 0.8 ] {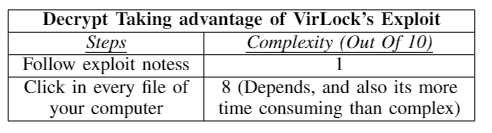}
	\caption{Decryption taking advantage of VirLock's exploit.}
	\label{fig:decrypt}
\end{figure}

\begin{figure}[H]
	\centering
	\includegraphics [ scale = 0.7] {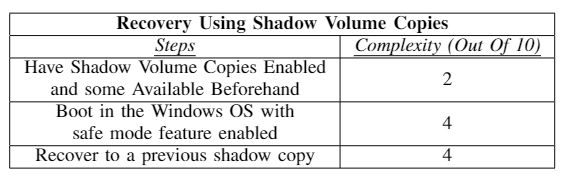}
	\caption{Recovery using shadow volume copies.}
	\label{fig:recovery}
\end{figure}

\begin{figure}[H]
	\centering
	\includegraphics [ scale = 0.75 ] {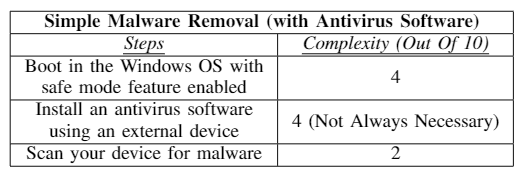}
	\caption{Simple malware removal (with antivirus software).}
	\label{fig:mremoval}
\end{figure}

\begin{figure}[H]
	\centering
	\includegraphics [ scale = 0.65 ] {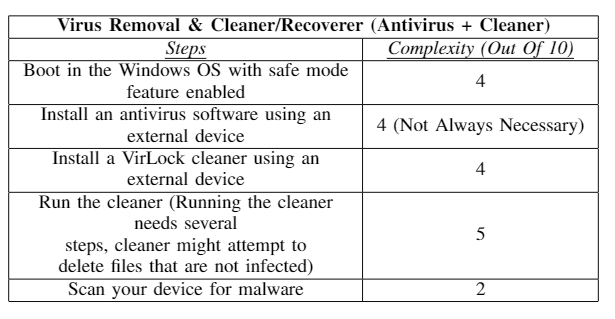}
	\caption{Virus removal \& cleaner/recoverer (antivirus \& cleaner).}
	\label{fig:vremoval}
\end{figure}

The \emph{complexity} variable has a range from 0 to 10 and is based on how complex it would be for an average computer user to perform one of those actions. The \emph{effectiveness} variable can take 3 different values (either low, medium or high) and it is based on the success rate that this technique has, as well as the percentage of the recovered files (in case where the entirety of them is not recovered). The risk of \emph{re-infection} variable takes the same values with the effectiveness variable and indicates whether the user is highly susceptible to get infected again or not, while using a certain recovery strategy. The strategies are also broken down into steps and a complexity value is also picked for each step with an average computer user in mind.

These tables (as well as the ransom payment payoff matrix) will help users infected with the VirLock malware adopt an optimal strategy for their infection scenario. The tables might also be useful to users with infections from similar malware. Having a well-thought-out plan (like those tables offer) beforehand can offer the user a clear advantage.

The effectiveness of the recovery using shadow volumes strategy is high, but this depends on whether the user has kept any of these copies beforehand and how old these copies are. The ``click on every file of your computer'' step depends on how many files the users stores in the computer. Installing an antivirus or anti-malware software is not always required, as one might have already been installed beforehand.

\section{Conclusion and further work} \label{sec:Conclusions}

Computer viruses can be classified into several categories depending on their characteristics. However, their similarity in relation to biological viruses is quite evident in many properties, as long as one correlates the appropriate viruses with each other, depending on their common behavioral elements. Studying them and the correlation between computer viruses and biological viruses offers an alternative look and approach on how to deal with both biological and computer viruses. It would also be interesting to identify unique strategies of one group and try to apply or simulate them in the other group, enforcing the appropriate parallelism, in order for new insights and solutions to arise.

\bibliographystyle{IEEEtranN}
\bibliography{CorrelationBiologicalComputerViruses}
\end{document}